\documentclass[aps,pra,preprint,amsmath,amssymb,showpacs]{revtex4}
\usepackage{graphicx}
\usepackage{dcolumn}
\usepackage{bm}

\newcommand{\eps}{{\bf \varepsilon}}
\newcommand{\sixj}[6]{\left\{\begin{array}{ccc}
#1 & #2 & #3 \\
#4 & #5 & #6
\end{array}\right\}}

\begin{document}

\title{Collisional shift and broadening of the transition lines in pionic helium}

\author{Boyan Obreshkov}

\author{Dimitar Bakalov}

\affiliation{Institute for Nuclear Research and Nuclear Energy,
Bulgarian Academy of Sciences, Tsarigradsko chausse\'{e} 72, Sofia
1784, Bulgaria}

\begin{abstract}

We calculate the density shift and broadening of selected dipole
transition lines of pionic helium in gaseous helium at low
temperatures  up to $T=12$ K and pressure up to a few bar. In the
approximation of binary collisions the shift and broadening depend
linearly on the density; we evaluate the slope of this linear
dependence for a few spectral lines of known experimental
interest, and also investigate
its temperature dependence. %
 We find a blue shift of the resonance frequencies of the
$(n,l)=(16,15) \rightarrow (16,14) $, $(17,16) \rightarrow
(17,15)$, and $(16,15)\rightarrow(17,14)$ unfavored transitions,
 and a red shift for the favored one $(17,16) \rightarrow
(16,15)$. The results are intended to significantly increase the
efficiency of the laser spectroscopy investigations of pionic
helium and help the interpretation of the experimental data.

\end{abstract}

\maketitle

\section{Introduction}

Pionic helium is a three-body system composed of a helium nucleus,
an electron in a ground state and a $\pi^-$ in a highly excited
Rydberg state with principal quantum number $n \sim (m^*/m_e)^{1/2}$, where $m^*$ is
the reduced mass of $\pi^-$ and the helium nucleus.
These states promptly de-excite via Auger transitions to
lower lying states which have large overlap with the helium
nucleus and subsequently undergo fast nuclear absorption for times
less than $10^{-12}$ s. However long-lived $\pi^-$ were observed in bubble-chamber experiments \cite{LL_pi}.
To explain this anomaly, Condo suggested that metastable atomic states of $\pi^-$
are formed
\begin{equation}
{\rm He}+ \pi^- \rightarrow [{\rm He}^+ \pi^-]_{nl} + e^-
\end{equation}
in which $\pi^-$ occupies states with high angular momentum $l \sim n-1$.
These Rydberg states are expected to retain nanosecond-scale lifetimes against
nuclear absorption and the electroweak decay $\pi^- \rightarrow \mu^-+\bar{\nu}_{\mu}$.
This is because for nearly circular orbits of $\pi^-$, the overlap with the helium nucleus is minimized,
whereas radiative de-excitation and the decay via Auger transitions
which lead to fast nuclear absorption are strongly supressed.
This hypothesis has been confirmed at TRIUMF in experiments with $\pi^-$ stopped in liquid helium \cite{nakamura_pi}.

A method for laser spectroscopy of metastable pionic helium atoms in gaseous helium
has been proposed \cite{Hori1,Hori2}. When comparing experimental
transition frequencies to three-body QED calculations of pionic
helium, the $\pi^-$ mass  can be determined with a fractional
precision better than $10^{-6}$. However systematic effects such
as collision-induced shift and broadening (S\&B) of the transition
lines, as well as the quenching of the metastable states can
prevent the experiment from achieving this high precision. Thus
reliable theoretical calculation for the density-dependent shift
and width is needed for the extrapolation of transition
wavelengths at zero target density.

\section{Collisional shift and broadening of the transition lines}

\subsection{Prerequisites: the potential energy surface}

The collisional shift and broadening of the laser stimulated
transition line $(n,l) \rightarrow (n',l')$ in pionic helium are
obtained in the impact approximation of the binary collision
theory of the spectral line shape \cite{Lindholm,Baranger,Allard}.
 This approach has already been applied in the
 calculations of the density effects on the line shape in
 antiprotonic helium \cite{prl2k,hfi11,long} and produced
 theoretical results in agreement with experiment \cite{torii};
 see also \cite{physrep} and references therein.
 The success of these calculations was due to the use of a
 highly accurate three-electron potential energy surface (PES) for
 the description of the binary interaction of an exotic
 helium atom with the atoms of the helium gas. The PES had been evaluated
 with {\em ab initio} quantum chemistry methods \cite{sapt} for
 nearly 400 configurations of the three heavy constituents of the
 interacting atoms (two helium nuclei and an antiproton or pion),
 selected to match the typical interpaticle distances in the
 experimentally interesting metastable states of exotic helium
 \cite{hfi14}.
 The configurations were parameterized with the length $r$ of the
 vector joining the heavy particles in the antiprotonic/pionic
 atom, the length $R$ of the vector joining its center-of-mass
 with the nucleus of the ordinary helium atom, and the angle
 $\theta$ between them.
 Subsequently, the numerical values of the PES
 at these $\sim$400 grid points were fitted with smooth functions
 $V(r,R,\theta)$.
 Two such fits, referred to as {\sc d47} and {\sc hn1}, have been widely used
 in calculations; they have been shown to produce numerical
 results for the S\&B in antiprotonic helium that differ by less
 than the overall numerical uncertainty.
 Earlier attempts to use the same PES in the evaluation of the
 S\&B of the spectral lines in pionic helium failed \cite{hfi14}
 because the typical distances $r$ between the pion and the helium
 nucleus in the pionic helium metastable states are
 outside the range for which the PES has been calculated, and the two
 fits {\sc d47} and {\sc hn1} produce wrong values when
 used for extrapolation of the PES. In the present work we carefully
 analyzed the behavior of these fits
 and established that the problem can be resolved by appropriately
 truncating the integration over $r$ in the expression for the
 effective state-dependent interatomic potentials in terms of
 $V(r,R,\theta)$ and the pionic helium atom wave function
 $\chi_{nl}(r)$
 \begin{equation}
 V_{nl}(R)=  \frac{1}{2} \int\limits_{0}^{\infty} dr \int d\theta \sin\theta
 \chi^2_{nl}(r) V(r,R,\theta)\ \rightarrow\
 V_{nl}(R)=  \frac{1}{2} \int\limits_{r_1}^{r_2} dr \int d\theta \sin\theta
 \chi^2_{nl}(r) V(r,R,\theta)\
 \label{veff}
 \end{equation}
 while keeping under control at each step the induced
 numerical uncertainties; the optimal values of the truncating
 parameters were found to be $r_1=0.2$ a.u., $r_2=1.3$ a.u.
 The S\&B of the four transition lines in
 pionic helium of experimental interest \cite{Hori1,private} were
 then evaluated by two alternative methods.
 Similar to the calculation of the S\&B in antiprotonic helium \cite{long},
 the obtained results
 differ by up to 30\% and should be regarded as boundaries of the
 intervals of uncertainty of the theoretical values of the density
 shift and broadening slopes in pionic helium.

 \subsection{Impact approximation}
 \label{barrang}

In this approximation, the slope of the density and
temperature-dependent collisional broadening $\Gamma_{fi}(T)$ and
shift $\omega_{fi}(T)$ are given by \cite{Baranger}
\begin{equation}
\alpha_{fi}(T)=\frac{\partial\Gamma_{fi}(T)}{\partial N}=\left
\langle \frac{\pi}{M k} \sum_{L=0}^{\infty} (2L+1) 2 \sin^2
\eta_{fi,L}(k) \right \rangle_T \label{barr1}
\end{equation}
and
\begin{equation}
\beta_{fi}(T)=\frac{\partial\omega_{fi}(T)}{\partial N}=-\left
\langle \frac{\pi}{M k} \sum_{L=0}^{\infty} (2L+1) \sin 2
\eta_{fi,L}(k) \right \rangle_T, \label{barr2}
\end{equation}
respectively, where $N$ denotes the number density of the target
gas, the labels $i$ and $f$ stand for the set of quantum numbers
of the initial and final pionic atom states $n,l$ and $n'l'$, and
$\langle \cdot \rangle_T$ denotes thermal average over the Maxwell
distribution. Both $\alpha$ and $\beta$ and expressed in terms of
the relative phase shifts
 \begin{equation}
   \eta_{fi,L}(k)=\delta_{iL}(k)-\delta_{fL}(k)
   \label{etafil}
 \end{equation}
 The partial
wave phases $\delta_L(k)=\delta_L(k,R \rightarrow \infty)$ are
obtained from the asymptotic solution of the variable phase
equation \cite{pam}
\begin{equation}
\frac{d}{dR} \delta_{nlL}(k,R)=-\frac{2M V_{nl}(R)}{k}  [\cos
\delta_{nlL}(k,R) j_L(kR)-\sin \delta_{nlL}(k,R)n_L(kR)]^2,
\end{equation}
subject to the boundary condition $\delta_{nlL}(k,0)=0$, where $k
=\sqrt{2 M E}$ is the wave-number of relative motion for a given
total collision energy $E$, $M$ is the reduced mass of the
collision system, $\{j_L(z),n_L(z)\}$ are the Riccati-Bessel
functions, and $V_{nl}(R)$ is defined in (\ref{veff}).

\begin{figure}
\begin{center}
\includegraphics[width=.9\textwidth]{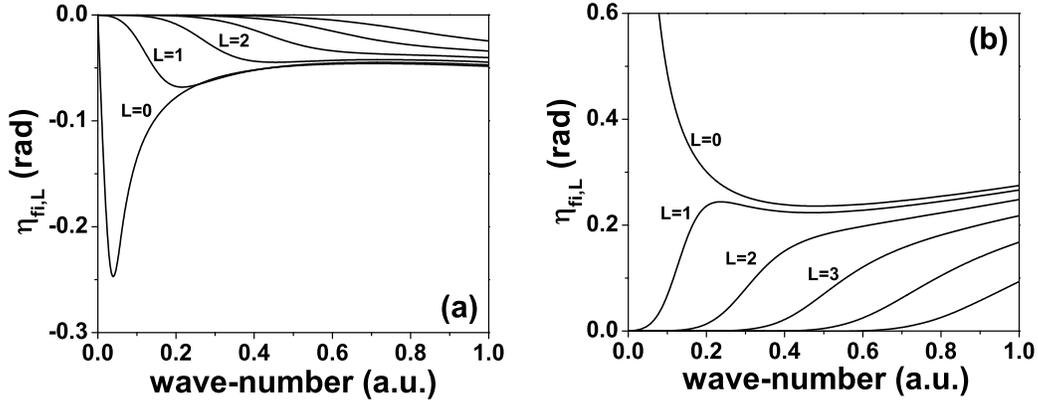}
\caption{The relative phase shifts $\eta_{fi,L}$ of
Eq.~(\ref{etafil}) for: (a) the unfavored laser-stimulated dipole
transition $(16,15) \rightarrow (16,14)$ in pionic helium in
gaseous helium medium, and (b) the favored dipole transition
$(17,16) \rightarrow (16,15)$. The different curves are labelled
by the orbital momentum $L$. } \label{fig:shifts}
\end{center}
\end{figure}

In Figs.~\ref{fig:shifts}(a-b) we show the scattering phase shifts
$\eta_L(k)$ for the "favored" transition $(17,16) \rightarrow
(16,15)$ and for the "unfavored" one $(16,15) \rightarrow
(16,14)$, respectively. For the unfavored transition in
Fig.~\ref{fig:shifts}(a), the scattering phases are negative and
are appreciably more than $-\pi/2$ over the entire range of wave
numbers, giving rise to blue shift of the transition frequency.
For $k < 0.5$, the scattering of $s,p$ and \mbox{$d$-waves} gives
dominant contribution to the dipole transition line shape. A
larger number of partial waves is required to converge the line
shift and width with the increased wave-number $k$. When $k \ge
1$, $\eta_L(k)$ exhibit linear dependence on the wave-number,
showing that in this regime the effective potentials act as
repulsive hard spheres with effective state-dependent radii, all
phase shifts would tend to zero in the high energy limit with $k
\gg 1$. Since the phase shifts are small $|\eta_L| \ll \pi/2$ over
the whole range of wave-numbers, the collisional broadening of the
transition line is negligible $\alpha \sim \sum_L (2L+1) \eta_L^2$
as compared to the line shift $\beta \sim -\sum_L (2L+1) \eta_L$.
In contrast, scattering phases are positive for the favored
transition shown in Fig.~\ref{fig:shifts}(b) resulting in
red-shift of the line center, because $\eta_L(k) < \pi/2$ for $k
> 0.1$ a.u. Since the \mbox{$s$-wave} scattering phase shift $\eta_0
\rightarrow \pi$ near  $k \rightarrow 0$, the effective potential
in the initial state $V_{17,16}(R)$ supports a single bound state
through the Levinson theorem. Since the contribution of
\mbox{$s$-wave} scattering becomes dominant towards threshold,
this bound state dramatically affects the transition line shape at
very low speeds with $k < 0.1$ as the center frequency undergoes a
blue shift when $\eta_0
> \pi/2$. However the thermally-averaged shift and width are
weakly affected by the low-velocity tail in the Maxwell
distribution, for instance at $T=6 K^{\circ}$, $\beta \approx -7 $
$10^{-21}$ GHz~cm$^3$ and $\alpha \ge 1 $ $10^{-21}$ GHz~cm$^3$.

The thermally averaged shift and width of the spectral lines for
the selected transitions in pionic helium, obtained with the fits
{\sc hn1} and {\sc d47} are given in columns 4 and 5 of
Table~\ref{numres}, respectively. The temperature dependence of
the line profile is relatively weak in gaseous helium. At low
perturber density $N=10^{21}$ cm$^{-3}$,  we find a blue shift of
the resonance frequencies for the $(n,l)=(16,15) \rightarrow
(16,14) $, $(17,16) \rightarrow (17,15)$, and
$(16,15)\rightarrow(17,14)$ transitions $\beta=2.5$~GHz, $6$~GHz
and $18$~GHz, respectively. For the favored transition $(17,16)
\rightarrow (16,15)$, the line center is red-shifted by $8$~GHz.
Thus in the absence of shape resonances in the potential
scattering, the direction of the shift reflects the sign of the
difference potential $\Delta V(R)=V_i(R)-V_f(R)$. The comparison
of the corresponding linewidths in Table~\ref{numres} makes
evident that the spectral line for the unfavored transition
$(16,15) \rightarrow (16,14)$ is only weakly affected by
collisions since it is broadened by $0.1$~GHz, which makes it
suitable for spectroscopic measurements in pionic helium.

\begin{figure}
\begin{center}
\includegraphics[width=.8\textwidth]{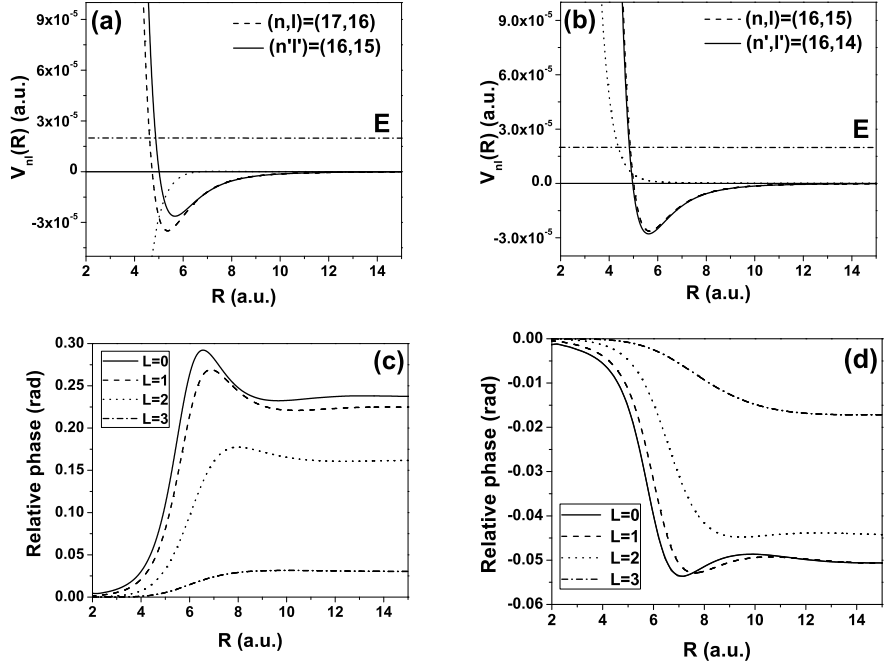}
 \caption{ Figs.(a-b) Potential energy curves $V_{nl}(R)$ for the
elastic scattering of pionic helium by an ordinary helium atom
prior to (dashed line) and after (solid line) the absorption of a
photon in the laser-stimulated dipole transitions $(16,15)
\rightarrow (16,14)$ and $(17,16) \rightarrow (16,15)$,
respectively. The potential energy difference $\Delta
V=V_{nl}-V_{n'l'}$ is given by a dotted line and the kinetic
energy $E$ of relative motion is indicated by the dashed-dotted
line. Figs.(c-d) Variable phase shifts (in radians) corresponding
to the potential energy curves in Figs.(a-b). The phase functions
are labelled by the orbital angular momentum $L$.
}\label{fig:phase}
\end{center}
\end{figure}

To further analyze the effect of the interaction energy
$V(r,R,\theta)$ in the ${\rm He}^+ \pi^-$-${\rm He}$ collision
system, in Figs.~\ref{fig:phase}(a-b) we plot the effective
state-dependent potentials for the transitions $(17,16)
\rightarrow (16,15)$ and $(16,15) \rightarrow (16,14)$,
respectively, together with the difference potentials $\Delta
V(R)$. Figs.~\ref{fig:phase}(c-d) present the corresponding
variable phase-functions at thermal collision energy with $T=6$~K.
For both transitions, the elastic scattering of $s$- and
\mbox{$p$-waves} gives the dominant contribution to the line shift
and width. The \mbox{$d$-wave} scattering is less efficient,
contributions of partial waves with $L \ge 3$ are suppressed due
to large centrifugal barrier. Figs.~\ref{fig:phase}(c-d) make
evident that a substantial part of the line shift comes from the
classically forbidden regions for relative motion with $4\le R \le
5$ a.u. The $s$- and \mbox{$p$-wave} variable phases rise steeply
in the classically allowed part of the scattering potentials $R >
5$ a.u., attain maxima near $R \approx 6$ a.u., then slightly fall
off and saturate in the asymptotic region with $R > 7$ a.u. Thus
the principal part of the line shift and width at thermal
collision energies are due to short-range binary encounters, the
effect of the long-range van-der-Waals tail $V(R) \sim C_6/R^6$
acts as a weak perturbation.

\subsection{Coupled partial waves}
\label{anisotr}

\begin{figure}
\begin{center}
\includegraphics[width=.8\textwidth]{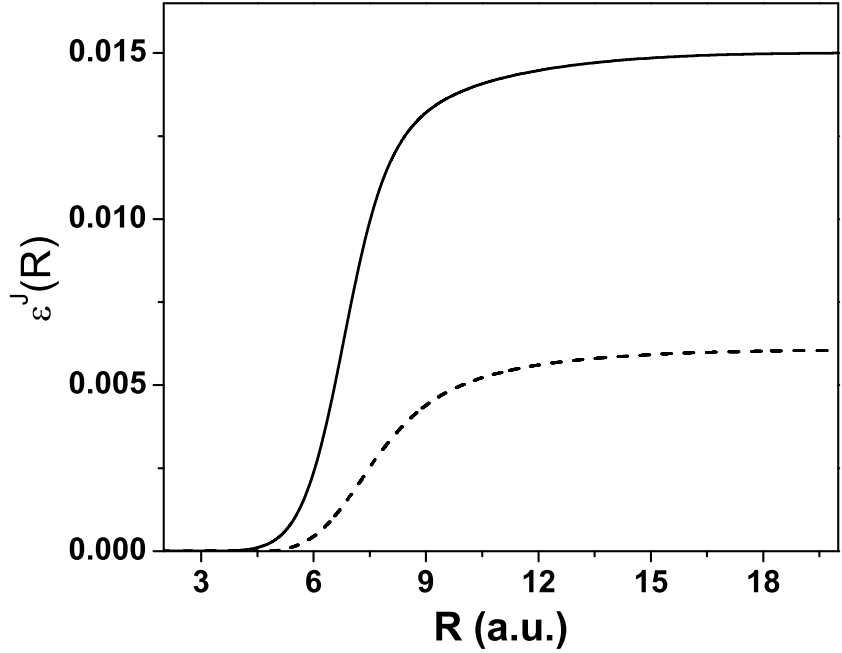}
 \caption{Collision-induced mixing of $s$- and
\mbox{$d$-waves} in the laser-stimulated dipole transition
$(17,16) \rightarrow (16,15)$ in pionic helium in gaseous helium
medium. The position-dependence of the mixing angle $\eps^J(R)$
prior to and after the photon absorption is given by the solid and
dashed lines, respectively. The total angular momentum is $J=16$
in the initial and $J=15$ in the final state. The collision energy
corresponds to $T=6 K^{\circ}$. }\label{fig:smat}
\end{center}
\end{figure}

As an independent test of our hypothesis that the dominant
contributions to the line shift and width are induced by effective
central state-dependent scattering potentials in both initial and
final states, we included the effect of the anisotropic part of
the interaction energy. The collision-induced shift and broadening
of the transition line shapes are evaluated from \cite{drake2006}
 \begin{eqnarray}
 &\alpha_{fi}+i \beta_{fi}=&\!\sum_{JJ'LL'} (-1)^{L+L'}
 (2J+1)(2J'+1)\sixj{J'}{J}{1}{l}{l'}{L}\sixj{J'}{J}{1}{l}{l'}{L'}
 \label{coupled}
 \\
 &&\times
 \left \langle \frac{\pi}{Mk} (\delta_{LL'}-S^J_{iL,iL'}(k)
 S^{J',\ast}_{fL,fL'}(k)) \right \rangle_T,
 \nonumber
 \end{eqnarray}
where $J$ and $J'$ are the total angular momenta before and after
the absorption of a photon.
 We obtain the  $S$-matrix elements ${\bf S}^J(k)$ from
 the asymptotic of the solutions
 of the coupled equations for the partial waves
 \begin{eqnarray}
 &&
 \frac{d{\bf S}^J(k,R)}{dR}= -i \frac{M}{k} [{\bf h}^{(2)}(kR)+
 {\bf S}^J(k,R)\cdot {\bf h}^{(1)}(kR)] {\bf V}^J(R)
 [{\bf h}^{(2)}(kR)+{\bf h}^{(1)}(kR) \cdot {\bf S}^J(k,R)],
 \nonumber
 \\
 &&
 {\bf S}^J(k)={\bf S}^J(k,R\rightarrow \infty),
 \end{eqnarray}
subject to the boundary condition ${\bf S}^J(k,0)={\bf I}$, where
${\bf I}$ is the unit matrix, ${\bf h}^{(1)}(z)$ and ${\bf
h}^{(2)}(z)$ are diagonal matrices of the Riccati-Hankel
functions. We approximate the matrix representation of the
interaction energy $V(\theta)=\sum_\lambda V_{\lambda}
P_{\lambda}(\cos \theta)$ by truncating its Fourier expansion over
Legendre polynomials at $\lambda=2$, i.e.
\begin{equation}
V^J_{nlL,nlL'}(R) \approx V_{nl}(R) \triangle(JlL) \delta_{LL'} +
V^{(2)}_{nl}(R) \sqrt{(2l+1)(2L+1)} (-1)^{l+L'-J}C^{l0}_{l0,20}
C^{L'0}_{L0,20} \sixj{2}{L}{L'}{J}{l}{l},
\end{equation}
where $C^{c \gamma}_{a \alpha, b \beta}$ are the Clebsch-Gordan
coefficients, $\triangle(abc)$ imposes the triangle condition and
\begin{equation}
V^{(2)}_{nl}(R)= \frac{5}{2} \int_{r_1}^{r_2} dr \chi^2_{nl}(r)
\int d \theta \sin \theta V(r,R,\theta) P_2(\cos\theta).
\end{equation}
At low-collision energies, the $S$-matrix for the coupled  $s$-
and $d$- waves can be parameterized in terms of two variable
phases $\delta^J_0$ and $\delta^J_2$ and a mixing angle $\eps^J$
\begin{equation}
{\bf S}^J(k,R) \approx \left(\begin{array}{cc}
e^{2 i \delta^J_0(k,R)} \cos 2 \eps^J(k,R) & i e^{i (\delta^J_0(k,R)+\delta^J_2(k,R))} \sin 2 \eps^J(k,R)  \\
i e^{i (\delta^J_0(k,R)+\delta^J_2(k,R))} \sin 2 \eps^J(k,R) &
e^{2 i \delta^J_2(k,R)} \cos 2 \eps^J(k,R)
\end{array}\right).
\end{equation}
In the $\eps^J \rightarrow 0$ limit, the $S$-matrix is diagonal
and the usual decoupled result is recovered. In
Fig.~\ref{fig:smat} we plot the  mixing angle $\eps^J$ as a
function of the separation $R$ between the colliding atoms at
thermal energy with $T=6$ K.

 The principal effect of the weak quadrupole coupling
 is to facilitate scattering of \mbox{$d$-waves} due to \mbox{$s$-wave} admixture in both initial
and final state wave-functions. The enhancement of the
\mbox{$d$-wave} elastic scattering is most effective in the
classically allowed region for both initial and final state
motions, where the colliding atoms experience the attractive part
of the scattering potential. The numerical results for the line
shape obtained with the {\sc hn1} fit are given in column 3 of
Table~\ref{numres}. As the comparison demonstrates, the anistropic
pairwise interaction has minor effect on the line shifts, which
justifies our approximation of using central state-dependent
potentials to describe the collision. However the line widths for
certain transitions are noticeably affected by the quadrupole
interaction; for example, the enhancement of \mbox{$d$-wave}
scattering causes additional broadening of the transition line
$(16,15)\rightarrow (17,14)$ by 1~GHz at low-temperatures. In
contrast the narrow line of the transition $(16,15) \rightarrow
(16,14)$ is insensitive to anisotropic interatomic interactions.

\subsection{Anderson's semiclassical method}
\label{anderson}

 The slopes of the shift and broadening of the spectral line of
 the transition can alternatively be represented in the form
 \cite{prl2k}

 \begin{eqnarray}
   &&
   \alpha_{fi}(T)=\frac{\partial\Gamma_{fi}}{\partial N}={\mathrm Re}\,\Phi,\
   \beta_{fi}(T)=\frac{\partial\omega_{fi}}{\partial N}={\mathrm Im}\,\Phi,
   \nonumber
   \\
   &&
   \Phi=\left\langle 2\pi\,v_T \int db\,b\,\left[
   1-\exp\left(-i\int dt\,
   (V_{n'l'}[|{\mathbf R}(t)|]-V_{nl}[|{\mathbf R}(t)|])
   \right)\right]\right\rangle_T,
   \label{anders}
 \end{eqnarray}
 where ${\mathbf R}(t)$ is the trajectory of the {\em classical}
 relative motion of the pionic and ordinary helium atoms (i.e. the
 vector joining the positions of the atoms at time $t$), while $b$
 and $v_T=\sqrt{2kT/M}$ are the impact parameter and the initial velocity
 of the relative thermal motion of the colliding atoms outside the
 interatomic potential range.
 Eq.~(\ref{anders}) in its initial form involving rectilinear
 trajectories ${\mathbf R}(t)={\mathbf R}_0+t{\mathbf v}_T$ was
 proposed by P.W.~Anderson \cite{anderson}; however, it fails to give
 reasonable estimates of the density effects at low temperature.
 The substantial modification that makes
 Eq.~(\ref{anders}) applicable in this case is to use instead the
 curvilinear trajectories determined by the binary interaction
 potential in the {\em initial state} for the transition of
 interest, $V_{nl}(R)$, as suggested in \cite{prl2k,long}. The
 numerical results for the density shift and broadening slopes
 for the transition lines in pionic helium of experimental
 interest, obtained with the fits {\sc hn1} and {\sc d47},
  are given in columns 6 and 7 of Table~\ref{numres},
 respectively.

 In agreement with the observations in subsection \ref{barrang},
 the detailed look at the multiple integral in Eq.~(\ref{anders}) reveals that
 the dominant contribution to $\Phi$ (through the integrals of the
 difference $V_{n'l'}(R(t))-V_{nl}(R(t))$ along the various classical
 trajectories) comes from the range $4\le R\le10$ a.u. For the
 favored transition up to 95\% of the value of $\Phi$ is
 accumulated in the range $4\le R\le6$ a.u., while for the
 unfavored transition the contributions from the domains
 $4\le R\le6$ and $6\le R\le10$ are balanced. This emphasizes once
 more the importance of using a realistic and accurate PES in the
 evaluation of the density effects; estimates based on the
 asymptotic shape of the interaction potentials, valid for $R>10$
 a.u., cannot be reliable.

 \section{Numerical results and discussion}

 The cumulative Table~\ref{numres} presents the numerical results
 on the slopes of the density shift and broadening,
 $\beta_{fi}(T)$ and $\alpha_{fi}(T)$, of four transition lines in
 pionic helium of declared experimental interest, obtained with
 the theoretical methods outlined in the preceding section and
 using two different fits of the PES. We see that in some cases
 the numerical values differ quite significantly and require
 further comments.

 \begin{enumerate}

 \item The two different fits {\sc hn1} and {\sc d47} produce
 values that differ by up to 20\%. As pointed out, this is related to
 the extrapolation of the fits outside the range of configurations
 relevant in antiprotonic helium for which the PES was initially
 evaluated. As long as both fits give close results for
 antiprotonic helium, we have no reasons to consider any of them
 as preferable; instead, the difference between the two values should be
 regarded as numerical uncertainty of the results that could only be
 eliminated with a new calculation of the PES for a wider range of
 configurations.

 \item The two theoretical approaches of subsections \ref{barrang}
 and \ref{anderson} produce results that differ by only a few percent for
 the favored transition $(17,16)\rightarrow(16,15)$ and by
 up to one third for the unfavored ones. The good
 agreement with experiment of the results for antiprotonic helium
 based on the use of classical trajectories, is not an argument to
 consider this approach as more credible in the case of pionic
 helium. We should therefore refer to these differences as a
 theoretical uncertainty of the results.

 \item Accounting for the contribution of the anisotropic part of
 the PES only slightly affects the shift slope $\beta_{fi}(T)$ and
 has a larger impact on the broadening slope $\alpha_{fi}(T)$
 which is still much smaller than the theoretical uncertainty.

 \item The moderate temperature dependence of the results
 is uniformly reproduced in all the calculations.
 \end{enumerate}

 \begin{table}[h]
 \caption{Slope of the density shift and broadening,
 $\beta_{fi}(T)$ and $\alpha_{fi}(T)$, for selected
 transition lines in pionic helium and temperatures $T$ in the
 range $4--12$~K, in units $10^{-21}$ GHz~cm$^3$.
 Listed are the numerical values obtained with Baranger's method
 with account of the anisotropic part of the potential energy
 surface (Eq.~(\ref{coupled})) and in the approximation of central
 interatomic potentials
 (Eqs.~(\ref{barr1},\ref{barr2})), as well as with Anderson's method
 (Eq.~(\ref{anders})), using either fit {\sc hn1} and {\sc d47}
 of the potential energy surface. The favored transition
 $(17,16)\rightarrow(16,15)$ is red-shifted, while all the
 unfavored transitions undergo a blue density-dependent shift.}
 \label{numres}
 \begin{tabular}{|@{\hspace{3mm}}c@{\hspace{3mm}}|@{\hspace{3mm}}r@{\hspace{3mm}}|
 @{\hspace{3mm}}r
 @{\hspace{3mm}}r@{\hspace{6mm}}r@{\hspace{6mm}}r@{\hspace{6mm}}r@{\hspace{3mm}}|}
 \hline
 \vrule height15pt width 0pt
 & & Eq.~(\ref{coupled}) & \multicolumn{2}{c}{Eqs.~(\ref{barr1},\ref{barr2})} &
 \multicolumn{2}{c|}{Eq.~(\ref{anders})}
 \\
 Transition & T, K$^{\circ}$ & {\sc hn1} & {\sc hn1} & {\sc d47}
 & {\sc hn1} & {\sc d47} \\
 \hline
 $(17,16)\rightarrow(16,15)$ &
   4 & $-$7.86(1.74) & $-$7.75(1.67) & $-$7.45(1.55) & $-$8.33(1.68) & $-$8.02(1.56) \\
 & 6 & $-$7.95(1.63) & $-$7.84(1.52) & $-$7.55(1.41) & $-$8.61(1.74) & $-$8.30(1.62) \\
 & 8 & $-$8.15(1.64) & $-$8.06(1.53) & $-$7.78(1.43) & $-$8.93(1.81) & $-$8.58(1.69) \\
 & 10& $-$8.35(1.68) & $-$8.28(1.57) & $-$8.00(1.47) & $-$9.18(1.88) & $-$8.84(1.75) \\
 & 12& $-$8.57(1.72) & $-$8.50(1.62) & $-$8.23(1.52) & $-$9.48(1.95) & $-$9.10(1.82) \\
 \hline
 $(17,16)\rightarrow(17,15)$ &
   4 & 6.48(0.85) & 6.27(0.75) & 6.41(0.78) & 4.43(0.36) & 4.70(0.40) \\
 & 6 & 6.35(0.77) & 6.16(0.66) & 6.00(0.68) & 4.52(0.34) & 4.78(0.37) \\
 & 8 & 6.20(0.73) & 6.04(0.63) & 6.16(0.64) & 4.62(0.33) & 4.83(0.35) \\
 &10 & 6.04(0.66) & 5.92(0.58) & 6.04(0.59) & 4.65(0.31) & 4.88(0.33) \\
 &12 & 5.93(0.60) & 5.82(0.53) & 5.94(0.54) & 4.75(0.32) & 4.93(0.32) \\
 \hline
 $(16,15)\rightarrow(16,14)$ &
   4 & 2.56(0.13) & 2.53(0.12) & 2.97(0.16) & 2.07(0.08) & 2.54(0.11) \\
 & 6 & 2.53(0.11) & 2.54(0.10) & 2.93(0.14) & 2.11(0.07) & 2.58(0.10) \\
 & 8 & 2.54(0.10) & 2.52(0.10) & 2.92(0.13) & 2.16(0.07) & 2.63(0.10) \\
 &10 & 2.56(0.10) & 2.51(0.09) & 2.93(0.12) & 2.21(0.07) & 2.66(0.09) \\
 &12 & 2.58(0.10) & 2.53(0.09) & 2.94(0.11) & 2.24(0.30) & 2.69(0.09) \\
 \hline
 $(16,15)\rightarrow(17,14)$ &
   4 & 17.84(8.91) & 17.69(7.94) & 17.60(7.74) & 11.37(2.66) & 11.82(2.84) \\
 & 6 & 18.04(8.38) & 18.10(7.41) & 18.01(7.19) & 11.74(2.63) & 12.19(2.78) \\
 & 8 & 17.90(7.97) & 18.06(7.22) & 17.99(6.98) & 12.11(2.64) & 12.56(2.78) \\
 &10 & 17.91(7.52) & 18.04(6.92) & 17.99(6.68) & 12.45(2.67) & 12.84(2.76) \\
 &12 & 18.08(7.18) & 18.17(6.67) & 18.12(6.42) & 12.70(2.68) & 13.11(2.76) \\

 \hline
 \end{tabular}
 \end{table}

 \section{Conclusion}

We have calculated the density shift and broadening of four dipole
transition lines in pionic helium in gaseous helium. At thermal
collision energies, we find a blue shift of the line center in the
unfavored transitions $(n,l)=(16,15) \rightarrow (16,14) $,
$(17,16) \rightarrow (17,15)$, and $(16,15) \rightarrow (17,14)$,
while the transition frequency is red-shifted for the favored
transition $(17,16) \rightarrow (16,15)$. The narrow collisional
line width ($0.1$~GHz) of the laser-induced resonance transition
$(n,l)=(16,15) \rightarrow (16,14)$ makes it suitable for
precision spectroscopy of pionic helium atoms. We demonstrate that
the major part of the collisional shift and width of the spectral
lines is induced by the short-range part of the interatomic
potential. Our result may be helpful in the extrapolation of the
transition wavelengths in pionic helium to zero density of the
perturbing helium gas.
 While the significant overall uncertainty of the S\&B
 slopes, discussed above, may not allow for a direct
 extrapolation of the experimentally observed resonance
 frequencies of the laser-induced transitions to
 zero target gas density, we expect that knowing
 the density shift to $\sim20$\%
 fractional accuracy will greatly enhance the efficiency of the
 precision laser spectroscopy study of pionic helium.

\end{document}